# the Galaxy

By J. K A L U Z N Y[1], W. K R Z E M I N S K I[2] AND B. M A Z U R[3]

[1]Warsaw University Observatory, Al. Ujazdowskie 4, 00–478 Warsaw, Poland

[2]Las Campanas Observatory, Casilla 601, La Serena, Chile

[3]Copernicus Astronomical Center, Bartycka 18, 00-716 Warsaw, Poland

## 1. Introduction

During the past few years we conducted a photometric survey of a large sample of rich and distant open clusters. Our goals included: a) identification of the oldest open clusters which can set an interesting lower limit to the age of the galactic disk; b) study of the chemical evolution of the Galaxy (metallicity vs. age and galactocentric distance). Multicolor CCD photometry was collected for about two dozen clusters. The observational data were obtained at KPNO and LCO observatories. In this contribution we present results for some most interesting objects from our sample.

## 2. The metal-rich clusters: Cr 261, BH 176 and Be 54

It is known that a substantial fraction of stars from the galactic bulge are objects of metallicity significantly higher than solar (Frogel & Whitford 1987). Recently, a few globular clusters were identified with $[Fe/H]$ likely to be solar or even higher then solar (eg. Bica *et al.* 1994). A signature of high metallicity is the presence of extremely red giants with $V - I > 3$ on the extension of the red giant branch. Such red stars were recently discovered in the very old open cluster NGC 6791 by Garnavich *et al.* (1994). In figure 1, we present color-magnitude diagrams (hereafter CMD) for three open clusters: Cr 261, BH 176 and Be 54. These objects belong to the group of the oldest open clusters known. Their ages can be estimated at 5-8 Gyr. Extremely red stars which are likely to be red giants are present in central parts of these clusters. CMD's of Cr 261 and Be 54 were published recently by Phelps *et al.* (1994). Our photometry of these objects is however deeper and more extended in comparison with their data. BH 176 is usually listed with globular clusters. Our photometry shows that its age is close to the age of NGC 6791 (7-9 Gyr). BH 176 can be classified either as a young globular cluster or as an old open cluster.

## 3. Very old open clusters Be 18 and Tr 5

In figure 2, we present the CMD's for two poorly known open clusters Be 18 and Tr 5. Their ages can be estimated at about 4 Gyr. A paper with detailed analysis of photometric data for these and several other clusters will be submitted by the end of 1994 to *Astronomy and Astrophysics*. Photometry for old clusters Be 17, Be 22, Be 29 and Be 54 was published recently by Kaluzny (1994, 1994a). Photometry for Cr 261 will be published soon by Mazur *et al.* (1995).

JK trip to Granada was supported by the KBN grant PB30400506.

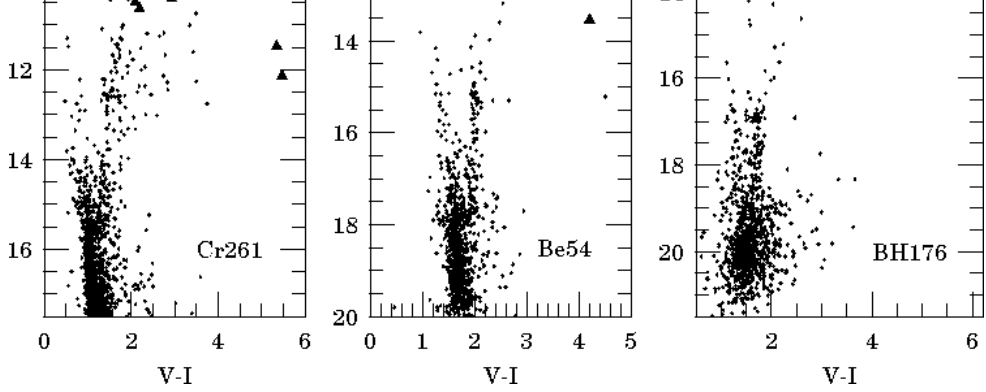

FIGURE 1. The $I$ vs. $V - I$ CMDs of Cr 261, Be 54 and BH 176. The large symbols mark extreme RGB candidates.

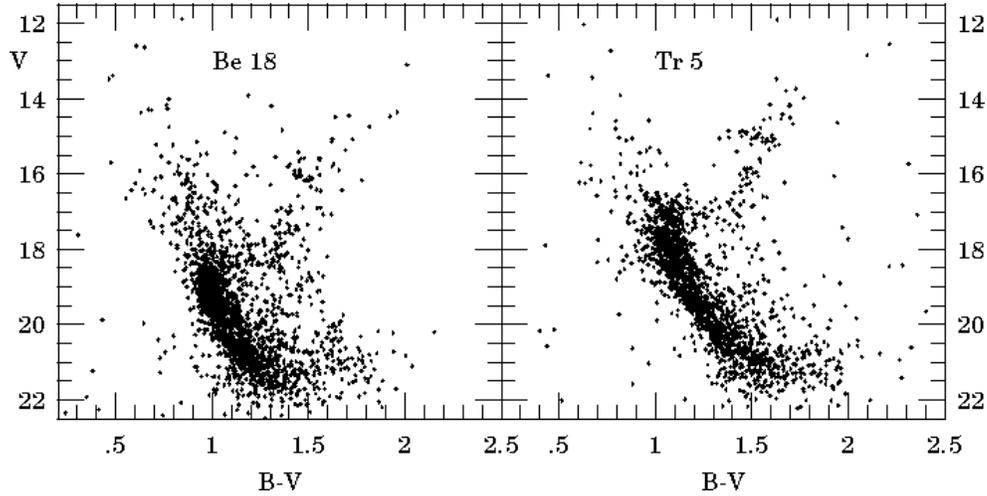

FIGURE 2. The CMD's of the old open clusters Be 18 and Tr 5.